\documentclass{article}
\usepackage{graphicx}

\title{Effects of ion drift on time-over-threshold charge
estimates using first-order integrators in thermal neutron MWPC.}

\author{Patrick Van Esch \\
Thomas Durand}

\begin{document}

\maketitle

\begin{abstract}
Time-over-threshold measurements of the output of a first order
integration amplifier can be used as a cheap way to estimate the
charge deposited on a wire in an MWPC.  Even if the time scale of
the first order system is seemingly much larger than the signal
development time, we noticed experimentally a significant
deviation from the relationship of time-over-threshold and the
charge, as would be naively expected from the exponential response
of the amplifier. We strongly suspect the current induced by the
residual ion movement to be at least partially responsible for
this observation.  It is of course well-known that the ion
movement is responsible for the bulk of the signal generation in a
wire chamber, this is not the point ; however, the surprise is
rather that this movement has influences on time scales which are
so long that the finite ion movement could be thought negligible.
In this paper we will treat theoretically the influence of ion
movement and of AC coupling on the expected time-over-threshold
relationship and confront this to the measurements on a small test
MWPC in thermal neutron detection.
\end{abstract}

\section{Introduction.}

In thermal neutron detection in gas detectors, a relatively short
trace of charge is generated as a result of the nuclear
interaction between the thermal neutron and an active component in
the gas such as He-3.  If this charge trace is detected in a MWPC,
it can be that several anode wires are hit.  When treating the
individual wire signals by pure discrimination, some information
gets lost and deteriorates the accuracy of the position
measurement. Although accurate charge measurements on each wire
are possible, this is sometimes considered as an overkill. A "poor
man's ADC", such as a measure of time-over-threshold can maybe
convey enough information in order to improve the accuracy of the
measurement, while keeping the simplicity of a discrimination
signal.  The relationship of collected charge versus
time-over-threshold (t-o-t) is deduced by the authors of
(\cite{becker} and \cite{manfredi}) purely based on the simulation
of the electronic part of the signal chain, which is probably
justified because they use a fast detector. However, in this paper
we will try to show that the relationship between the t-o-t and
the charge in a MWPC cannot just be extracted from the impulse
response of the amplifier alone. For a first-order amplifier
transfer function, the t-o-t is significantly influenced by the
ion movement that induces the current signal in the anode wire.
Also AC-coupling has a strong influence on this t-o-t measurement.

\section{Experimental setup and electrostatics.}

A small prismatic MWPC has been build where 16 anode wires of 25
micron diameter and 16 cathode wires of 75 micron diameter form
two parallel planes, at 2 resp. 6 mm from the rear of the detector
wall.  All wires are parallel and are spaced by 2.5 mm in each
plane. The total detector gap is 36 mm, and a negative potential
can be applied to the front window of the detector.  The detector
is filled with 1 bar of CF4 and 100 mbar of He-3.  The anodes are
polarized at 1750 V and the cathodes are put to ground potential,
as are the walls of the detector except for the front window,
which is put to -500V.

The calculation of the potential can be done analytically in this
prismatic set up, if one uses the conformal mapping (with $F$ the
elliptic integral of the first kind\footnote{$F(x,k^2) = \int_0^x
\frac{d \psi}{\sqrt{1-k^2 \sin^2 \psi}}$, see p 43 of
\cite{jahnkeemde}} and $am$ its inverse function, the Jacobi
Amplitude):
\begin{equation}
w(z) = 21.7939 F(\arcsin(z),k^2 = 0.81561)
\end{equation}
and its inverse:
\begin{equation}
z(w) =  \sin( am(0.0458844 w, k^2 = 0.81561))
\end{equation}
which maps the rectangle in $w$ of width 100mm and height 36mm
describing the detector cross section into the upper half plane in
$z$, see \cite{schaum}.

The potential in the $z$ plane takes on the form \cite{appel},
\cite{jackson}:
\begin{eqnarray}
  V(z) &=& V_0 Re\left(2i/\pi
\ln\left(\frac{1+z+\sqrt{1-z^2}}{z-1-\sqrt{1-z^2}}\right)\right) \nonumber \\
   &-& \sum_{k=1}^{32}
\frac{q_k}{2\pi} (\ln{|z-z_k|}-\ln|z-\widetilde{z_k}|)
\end{eqnarray}
where the $q_k$ are the charges\footnote{We make here the
approximation that we can neglect the dipole and higher multipole
field contributions of the charge distribution on the surface of
the wire, which is justified by the small diameter of the wire.}
on the wires and $z_k = z(w_k)$ are their transformed positions in
the $z$-plane.  $V_0$ is the potential applied to the front
window. Imposing the potentials at the transformed wire radii then
solves for the charges, giving us the solution $V(z)$ in the
$z$-plane, and hence the correct electrostatic potential
$\phi(w)=V(z(w))$ in the detector volume.

The drift field (between the front window and the cathode plane)
then turns out to be equal to 216 V/cm, which gives rise to an
electron drift speed of about $4 cm/\mu s$ according to Magboltz
\cite{magboltz}. Given that in the gas mixture, the ionization
trace has a total length of about 5.6mm \cite{srim}, this means
that the longest charge deposition times (when the trace is
parallel to the drift field) are about 140ns.

\section{Experimental observation of d-t-o-t.}

We have a first order trans-impedance amplifier, with unit
response
\begin{equation}
\label{eq:firstorder}
 v_1(t) \sim e^{-t/\tau}
\end{equation}
$\tau$ being equal to 120ns in our case (this has been verified
experimentally).

We can expect, from the signal maximum onward, an exponentially
decreasing signal, assuming that, as long as there is an avalanche
going on, the signal will still rise (during a time which is less
than 140ns). There is then a very simple relationship between this
decrease-time-over-threshold (d-t-o-t) $T_e$ as a function of the
ratio $1/n$ of the threshold and the signal maximum:
\begin{equation}
\label{eq:expotot} T_e(n) = \tau \ln{n}
\end{equation}
It would be the same as the genuine time over threshold (t-o-t) if
the rise time were negligible (impulse response).  So $T_e(n)$ is
a \emph{theoretical} relationship between the ratio of the maximum
amplitude of a signal to a fixed threshold value, $n$, and a
duration, and this duration ought to be equal to the t-o-t of a
first order system when the rise time is negligible, or equal to
the d-t-o-t of a first order system, even when the rise time is
not negligible.  The t-o-t, in such a case, will then be the sum
of the d-t-o-t and the rise time.

\begin{figure}
  \includegraphics[width=8cm]{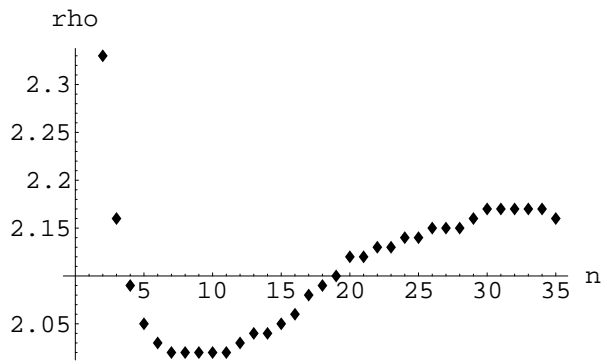}
 \caption{The experimentally measured ratio $\rho(n)$ of d-t-o-t
over the prediction of equation \ref{eq:expotot}.}
  \label{fig:ratiodata}
\end{figure}

Signals have been recorded using a digital oscilloscope with 8 bit
vertical resolution and a sample period of 5ns.  We will work with
the 'decrease time over threshold', which is the time measured
from the moment the signal reaches its maximum to the time it
falls below the threshold value. As said before, for a first order
system, t-o-t and d-t-o-t are equal when the rise time is
negligible. We now define the ratio $\rho$ of the experimentally
measured d-t-o-t to what we expect using equation
\ref{eq:expotot}, with $n = V_{max}/V_{thresh}$,
\begin{equation}
\rho(n) = \frac{\textrm{d-t-o-t}(n)}{T_e(n)}
\end{equation}
$\rho(n)$ measures the deviation from the exponential model; we
expect $\rho(n) = 1$ if the simple exponential model is correct.
Experimentally, we find the result displayed in figure
\ref{fig:ratiodata}. Clearly, some aspect of the signal is not
understood, because equation \ref{eq:expotot} seems to predict
times which are about a factor of 2 below the measured values. We
strongly suspect that the residual induced currents by ion
movement are at least partially responsible for this prolongation,
and we will work out their effect theoretically using a simple
model to verify whether we obtain the right order of magnitude
correction.

\section{The induced signal.}

\begin{figure}
  \includegraphics[width=8cm]{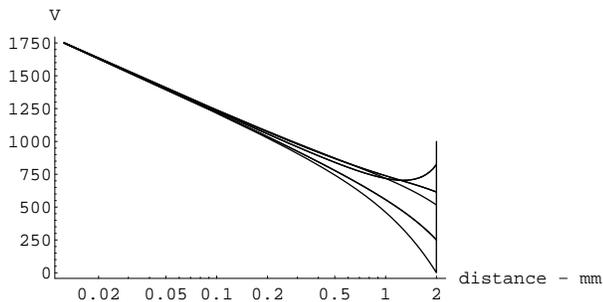}
  \caption{The applied potential around the 8-th anode wire,
  in 8 different radial directions.}
  \label{fig:driftpot}
\end{figure}

The electrostatic situation around wire 8 in our test chamber
allows us to establish the current induction from radial ion
movement. Looking at the applied potential around anode wire 8, we
see that up to about $200 \mu m$ we have a nice isotropic ln(r)
behavior as shown in figure \ref{fig:driftpot} ; after that, the
potential becomes more involved and direction-dependent.

\begin{figure}
  \includegraphics[width=8cm]{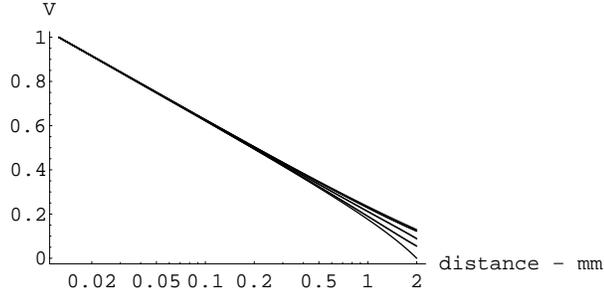}
  \caption{The sense potential around the 8-th anode wire, in 8 different
  radial directions.}
  \label{fig:sensepot}
\end{figure}

If, in order to apply the Shockley-Ramo theorem \cite{knoll}, we
bring the 8th anode wire to 1V and all the rest to 0V, we see in
figure \ref{fig:sensepot} that the sensing potential around that
anode goes in $\ln (r)$ to a significant distance from the wire
(at least $500 \mu m$).

So up to about 20 anode wire radii\footnote{In figure
\ref{fig:sensepot} it is also clear that at 20 anode wire radii,
the ions have induced about half of their charge.}, we can write,
for the applied electric field: $E_a(r) = E_a^0 R_{anode}/r$ and
for the sense field: $ E_s(r) = E_s^0 R_{anode}/r$, where $E_a^0$
and $E_s^0$ are respectively the applied electric field and the
sense field at the surface of the anode wire (and which are
proportional to the respective charges calculated on the wire).
For the applied field in the test detector, we find $E_a^0 = 200
kV/cm$.

Next we determine the ion movement in this field. Let us first
treat the low field case and assume constant ion mobility $\mu$.
In that case, the induced current in the anode wire is given by:
\begin{equation}
i(t) = \mu E_a(r) E_s(r)
\end{equation}
and we are in exactly the same situation as the standard signal
development in a cylindrical proportional counter, treated for
example in \cite{knoll}.  The solution is given by:
\begin{equation}
i(t) = \frac{\frac{1}{2}E_s^0 R_{anode}}{t+\frac{R_{anode}}{2
E_a^0 \mu}}
\end{equation}
We can define the "drift time constant" $t_0$ as:
\begin{equation}
t_0 = \frac{R_{anode}}{2 E_a^0 \mu}
\end{equation}
The problem the authors have for the CF4-He-3 gas mixture is that
no value for the ion mobility seems to be available in the
literature.  However, if we assume a mobility of the order of $1
cm^2/Vs$ which is a typical mobility for many gasses in n.c.
\cite{sauli} we find $t_0 \simeq 3 ns$. The amplifier time
constant (of the first order system) is 120 ns, so at first sight,
$t_0$ is about 2 orders of magnitude smaller than $\tau$ and can
be thought to be negligible.

If we treat the high field limit, which is probably more
appropriate for the actual field strengths, and where the ion
velocity is proportional to the square root of the field, we can
write:
\begin{equation}
i(t) = \alpha \sqrt{E_a(r)} E_s(r)
\end{equation}
Working out the differential equation, we find again the same
structure for the time dependence:
\begin{equation}
i(t) = \frac{ \frac{2}{3} E_s^0 R_{anode} }{ t + \frac{2
R_{anode}}{3 \sqrt{E_a^0}\alpha} }
\end{equation}
so we can again define a constant $t_0$:
\begin{equation}
t_0 =\frac{2 R_{anode}}{3 \sqrt{E_a^0}\alpha}
\end{equation}

The authors didn't find any indication of the value of this
constant in the literature.  We can make an educated guess in the
following way: $t_0$ in the high field limit over $t_0$ in the low
field limit is equal to $\frac{4}{3}\sqrt{\frac{E_a^0}{E^{*}}}$
with $ E^{*} $ the field strength which separates the high field
and the low field region (of the order of $50KV/cm$).  This then
leads to a high field $t_0$ value of the order of $8ns$.

The important point, however, is that no matter whether we work
with the high field or the low field approximation, the analytical
form of the induced current is the same.

\section{Electronic response with first-order system.}

In the previous section, we tried to argue that the impulse
response of the linear system "electron arrives in avalanche
region" to "current induced in anode wire" takes on the form:
\begin{equation}
i(t) = \frac{1}{t+t_0}
\end{equation}
and this both in the low field case and the high field case (but
with different values of $t_0$).

Using our first order electronic response as in equation
\ref{eq:firstorder} which converts this current signal into a
voltage signal, we obtain as an overall impulse response:
\begin{equation}
\label{eq:iondc}
 v(t) = e^{-\frac{t+t_0}{\tau}}\left(
Ei\left(\frac{t+t_0}{\tau}\right)-Ei\left(\frac{t_0}{\tau}\right)\right)
\end{equation}
with $Ei$ the exponential integral function\footnote{$Ei(x) = -
\int_{-x}^{\infty}\frac{e^t}{t}dt$, see p 925 of
\cite{gradshteyn}}. Often, there's an extra AC coupling present
(RC high pass filter) which takes on the form (in the Laplace
domain) $\frac{\theta s}{1+\theta s}$. Taking this into account,
the overall impulse response becomes:
\begin{eqnarray}
\label{eq:ionac}
  v_{\theta}(t) &=& \frac{\exp\left(-\frac{(t+t_0)
  (\tau+\theta)}{\tau \theta}\right)\theta}{\theta-\tau} \times
 \left[
  -e^{\frac{t+t_0}{\theta}}
  \theta
  \left(Ei(\frac{t_0}{\tau})-Ei(\frac{t+t_0}{\tau})\right)\right.
  \nonumber \\
   & &\left. +  e^{\frac{t+t_0}{\tau}}\tau \left(Ei(\frac{t_0}{\theta})-
   Ei(\frac{t+t_0}{\theta})\right) \right]
\end{eqnarray}

\section{Time-over-threshold in our model.}

\begin{figure}
  \includegraphics[width=8cm]{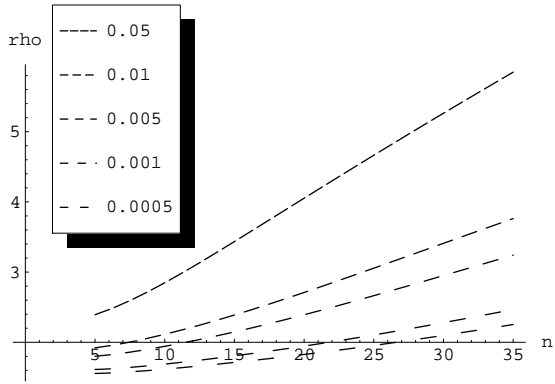}
  \caption{The ratio $\rho(n)$ using the model for ion movement,
  DC coupling, for different values of $t_0/\tau$.}
  \label{fig:ratiodc}
\end{figure}

We will now calculate the ratio $\rho(n)$ of the d-t-o-t of our
model given in equation \ref{eq:iondc} over the simple result of
equation \ref{eq:expotot}. The results are shown in figure
\ref{fig:ratiodc}, for different values of the ratio $t_0/\tau$.

\begin{figure}
  \includegraphics[width=8cm]{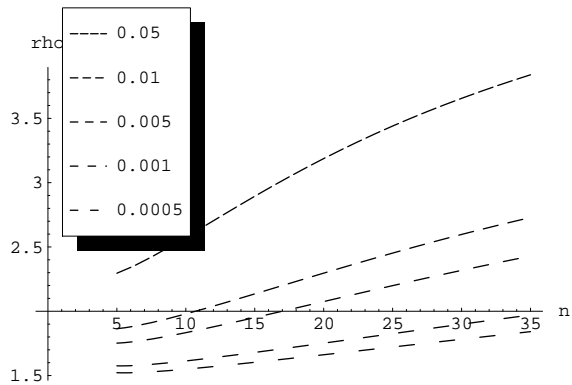}
  \caption{The ratio $\rho(n)$ using the model for ion movement,
  AC coupling, for different values of $t_0/\tau$.}
  \label{fig:ratioac}
\end{figure}

In our measurements, we had an amplifier which was AC coupled,
with a time constant $\theta \simeq 195.0 \tau$. This AC coupling
has a time constant which is more than two orders of magnitude
larger than $\tau$ and could be thought of being negligible. If we
use equation \ref{eq:ionac} with this value, we see in figure
\ref{fig:ratioac} that there is nevertheless a very strong
influence on the d-t-o-t due to this AC coupling as compared to DC
coupling in figure \ref{fig:ratiodc}.

Comparing these model predictions to our experimental results of
figure \ref{fig:ratiodata}, although one cannot talk about a fully
quantitative agreement\footnote{The fact that for small values of
$n$, the experimental curve rises is simply due to the finite
charge collection time (which can be of the order of 140 ns) which
lengthens the experimental d-t-o-t for small values of $n$ and
hence small values of time.}, the calculated values of $\rho$ are
in the right ballpark (around a factor of 2) for reasonable values
of the unknown parameter $t_0$.  This strongly suggests that the
ion movement has a non-negligible influence on the t-o-t
measurements.

\section{Conclusion}
Starting from the observation that a first order model of the
electronics makes predictions of t-o-t which are about a factor of
2 smaller than what is experimentally measured on a test MWPC, we
tried to explain the origin of this discrepancy with a simple
model of the current induced by the ion movement.  This comparison
indicates that the signal induced by residual ion movement has to
be taken into account when modeling t-o-t systems using a MWPC,
even when time scales of $\tau$ and $t_0$ differ by 2 orders of
magnitude. As a side result, it is also shown that an AC coupling,
even with a large time constant, also has a strong influence on
the t-o-t; in this case, the t-o-t values are shortened.

\end{document}